\documentclass[preprint,aps,amsmath,amssymb,12pt]{revtex4}

\usepackage{epsfig}
\usepackage{slashed}
\usepackage{graphicx}
\usepackage{multirow,color}
\usepackage{amsmath}
\usepackage{float}
\usepackage{diagbox}
\usepackage{CJK}
\usepackage{color}
\usepackage{xcolor}
\usepackage{times}
\usepackage{subfigure}
\usepackage{bm}
\usepackage{braket}
\usepackage{booktabs}
\usepackage{array}
\usepackage[mathscr]{euscript}
\usepackage{caption}
\usepackage[justification=centering]{caption}
\usepackage{makecell}

\makeatletter

\newcommand{\Rmnum}[1]{\expandafter\@slowromancap\romannumeral #1@}
\makeatother
\graphicspath{{fig/}}
\begin{document}
\title{Single production of singlet vector-like leptons at the ILC}
\author{Chong-Xing Yue$^{1,2}$}
\thanks{cxyue@lnnu.edu.cn}
\author{Yue-Qi Wang$^{1,2}$ \footnote{Corresponding author}}
\thanks{wyq13889702166@163.com}
\author{Xiao-Chen Sun$^{1,2}$}
\thanks{xcsun0315@163.com}
\author{Xin-Yang Li$^{1,2}$}
\thanks{lxy91108@163.com}

\affiliation{
$^1$Department of Physics, Liaoning Normal University, Dalian 116029, China\\
$^2$Center for Theoretical and Experimental High Energy Physics, Liaoning Normal University, China
}

\begin{abstract}

Vector-like leptons (VLLs) as one kind of interesting new particles can produce rich phenomenology at low- and high-energy experiments. In the framework of the singlet vector-like lepton with scalar (VLS) model, we investigate the discovery potential of VLL via its single production at the International Linear Collider (ILC) with the center of mass energy $\sqrt{s} =$ 1 TeV and the integrated luminosity $\mathcal{L}$ = 1 ab$^{-1}$, taking into account the appropriate polarization. For the signal and standard model (SM) background analysis, we have considered two kinds of decay channels for the W boson, i.e. pure leptonic and fully hadronic decay channels. Our analytic results show that the parameter space $M_{F}\in$ [300, 675] GeV and $\kappa \in$ [0.0294, 0.1] might be detected by the proposed ILC for pure leptonic decay channel. For fully hadronic decay channel, larger detection region of the parameter space are derived as $M_{F}\in$ [300, 700] GeV and $\kappa \in$ [0.0264, 0.0941].

\end{abstract}


\maketitle
\section{Introduction}
With the discovery of the Higgs boson by the a toroidal large hadron collider apparatus (ATLAS) and compact muon solenoid (CMS) collaborations at the Large Hadron Collider (LHC) in 2012~\cite{ATLAS:2012yve,CMS:2012qbp}, the standard model (SM) of particle physics has was further validated. So far, the SM is a remarkably successful theory capable of explaining a large number of observations. However, it is clear that the SM lacks some of the vital ingredients to help us understand nature, such as the candidates of dark matter~\cite{Bertone:2004pz,Feng:2010gw}, the excess of matter over antimatter~\cite{Planck:2013pxb}, gauge hierarchy~\cite{Feng:2013pwa}, neutrino masses~\cite{Gonzalez-Garcia:2007dlo}, as well as the fact that the masses and mixing patterns of the SM leptons cannot yet be clearly determined~\cite{Morais:2021ead}. Besides, there are also deviations between the experimental measurements and the SM predictions for certain physical observations. For instance, the measurement of the  muon anomalous magnetic moment (AMM) $a_{\mu }$ indicates that the significance of the discrepancy between the experimental average and the SM prediction  approaches to 4.2$\sigma$~\cite{Muong-2:2021ojo}. Therefore, we need new physical scenarios to provide a more natural solution to the above problems and  the existed anomalies.

Considering the flavor democracy (FD) hypothesis that may enlighten mass pattern of the SM fermions, the existence of extra fermions beyond the three generations of the SM is necessary. The Higgs data from LHC imposes strong constraints on the additional chiral fermions, while the non-chiral fermions, generally referred to as vector-like fermions, are much less constrained~\cite{Kuflik:2012ai}. Due to the vector-like nature of non-chiral fermions, they can evade constraints from fourth-generation searches~\cite{Eberhardt:2012gv}. Vector-like fermions with masses generated independently of the Higgs mechanism become increasingly attractive and appear in numerous new physics models beyond the SM, such as little Higgs models~\cite{Arkani-Hamed:2002ikv,Perelstein:2003wd,Carena:2006jx,Han:2003wu}, composite Higgs models~\cite{Anastasiou:2009rv,Kong:2011aa,Gillioz:2012se,Keren-Zur:2012buf,Falkowski:2013jya}, models of warped or universal extra-dimension~\cite{Agashe:2006wa,Contino:2006nn,Li:2012zy,Huang:2012kz,Redi:2013pga}, grand unified theories with non-minimal supersymmetric extensions~\cite{Kang:2007ib,Babu:2008ge,Martin:2009bg,Graham:2009gy}. Very recently, Ref.~\cite{Adhikary:2024esf} has reexamined theoretical constraints on the new physics models with vector-like fermions.

Vector-like fermions include vector-like quarks (VLQs) and vector-like leptons (VLLs), where VLLs are defined as colourless, spin $-1/2$ fermions. The left- and right-handed components of VLLs transform in the same way under the electroweak symmetry group. Like the SM leptons, VLLs have three generations and the same lepton numbers with their corresponding SM leptons. VLLs might explain the "Cabibbo angle anomaly"~\cite{Crivellin:2020ebi}, the muon AMM anomaly~\cite{Megias:2017dzd} and help obtain the realistic Higgs mass within the framework of the composite Higgs model~\cite{Carmona:2014iwa,Carmona:2015ena,Carmona:2017fsn}, while also being used to study the phenomenology of dark matter~\cite{Guedes:2021oqx}.

The current research status of VLLs can be discovered in Ref.~\cite{Baspehlivan:2022qet}. Although VLLs have the same status as VLQs from particle phenomenology, there are not many direct searches for VLLs. Prior to direct searches at the LHC, a lower bound on the VLLs mass $M_{VLL}>101.2$ GeV is placed by L3 collaboration at the large electron-positron collider (LEP)~\cite{L3:2001xsz}. Using the LHC data about multilepton events, collected in 2016-2018 corresponding to an integrated luminosity of  138 fb$^{-1}$, the CMS collaboration has given the constraints on the VLL masses, which excludes the doublet VLLs with mass below 1045 GeV at 95\%  confidence level (CL), while the singlet VLLs with mass in the range from 125 GeV to 150 GeV are excluded~\cite{CMS:2022nty}. The analytical results of the ATLAS collaboration show that  the doublet VLLs with mass in the range from 130 GeV to 900 GeV are excluded  at 95\% CL~\cite{ATLAS:2023sbu}. We believe that future runs of the LHC (LHC Run-3 and High Luminosity-LHC) will give more interesting results about searching for VLLs.

The vector-like lepton with scalar (VLS) models~\cite{Hiller:2020fbu,Hiller:2019mou,Hiller:2019tvg} satisfy asymptotically safety that tames the UV behavior towards the Planck scale~\cite{Litim:2014uca,Bond:2018oco,Bond:2017lnq,Bednyakov:2023fmc,Litim:2023tym} and are allowed to explain the discrepancies between the measured values of the electron and muon AMMs and the SM predicted values. The VLS models, which can be divided into the singlet and doublet VLS models, predict the existence of VLLs and new scalars. The couplings of these new particles with the SM  particles through renormalizable Yukawa interactions. The masses of VLLs can be as light as a few hundred GeV, so they can be probed at different colliders~\cite{L3:2001xsz,Bissmann:2020lge,Yue:2024sds,Das:2020gnt,Das:2020uer,Bhattacharya:2021ltd}.

It is well known that lepton colliders can offer a cleaner environment compared to hadron colliders and the polarized incident beams can improve the discovery potential of new particles. Thus, in this work, we will explore the possibility of detecting  VLLs at the International Linear Collider (ILC), taking into account the appropriate polarization for our analysis. Considering the lower limits on the masses of the singlet and doublet VLLs, we concentrate on the searches for VLL within the framework of the singlet VLS model through its single production with subsequent decays to the pure leptonic final state and the fully hadronic final state at the ILC with the
center of mass energy $\sqrt{s} =$ 1 TeV and the integrated luminosity $\mathcal{L}$ = 1 ab$^{-1}$~\cite{ILC:2019gyn,Bambade:2019fyw}.

This paper is organized as follows. In section II, we briefly present the main content of the singlet VLS model, which is related our numerical calculation. In section III, we offer a detailed analysis of the possible of detecting singlet VLL through its single production processes based on the ILC detector simulation. For the analysis of signal, we consider two kinds of decay modes for the W boson, i.e. pure leptonic and fully hadronic channels. Our conclusions are given in section IV.

\section{The Theory Framework}
  The VLS models ~\cite{Hiller:2020fbu,Hiller:2019mou,Hiller:2019tvg} predict three generations of VLLs, denoted as $\psi_{L,R}$ and each generation of VLLs is colorless with hypercharge $Y=-1$ and $-1/2$, which can be regarded as $SU(2)_L$ singlets and doublets, respectively. Considering the constraints on the masses of the doublet and singlet VLLs from the LHC~\cite{CMS:2022nty,ATLAS:2023sbu}, we will focus our attention on the possibility signal of the  singlet VLL via its single production at the 1 TeV ILC.

  The new Yukawa couplings in the framework of the singlet VLS model read~\cite{Hiller:2020fbu}
\begin{eqnarray}
\begin{split}
\mathcal{L}_{\text{Y}}^{\text{singlet}}
	=
    &-\kappa\bar{L}H\psi_{R} -\kappa'\bar{E}S^{\dagger }\psi_{L}-y\bar{\psi}_{L}S\psi_{R}+h.c.,
\end{split}
\end{eqnarray}
where E, L and H denote the SM  singlet, doublet leptons and Higgs boson, respectively. S represents new scalar introduced to the singlet VLS model that is singlet under the SM gauge groups. Due to the conservation of each lepton flavor meeting with $SU(3)$-flavor symmetry, the new Yukawa couplings $y, \kappa, \kappa'$ become single couplings instead of tensors.

After electroweak symmertry breaking, the couplings of the singlet VLL with the electroweak gauge bosons and new scalars are given by~\cite{Bissmann:2020lge,Hiller:2020fbu}
\begin{eqnarray}
\begin{split}
\mathcal{L}_{\text{int}}^{\text{singlet}}
	=
    &-e\bar{\psi}\gamma^{\mu}\psi A_{\mu}+\frac{g}{\cos\theta_{w}}\bar{\psi}\gamma^{\mu}\psi Z_{\mu} + (-\frac{\kappa}{\sqrt{2}}\bar{\ell }_{L}\psi_{R}h-\kappa'\bar{\ell }_{R}S^{\dagger }\psi_{L}+g_{S}\bar{\ell }_{R}S^{\dagger }\ell_{L}\\
    &+g_{Z}\bar{\ell }_{L}\gamma^{\mu}\psi_{L}Z_{\mu}+g_{W}\bar{\nu}\gamma^{\mu}\psi_{L}W_{\mu}^{+}+h.c.).
\end{split}
\end{eqnarray}
Here, $A_{\mu}$, $Z_{\mu}$ and $W_{\mu}$ denote photon, Z boson and W boson, respectively. $e$, $\theta_{w}$, $g$ correspond to the electromagnetic coupling, the weak mixing angle and the $SU(2)_L$ coupling. The coupling constants $ g_{S}$, $ g_{Z} $ and $ g_{W}$  are represented by the following formulas
\begin{eqnarray}
\begin{split}
&g_{S}=\frac{\kappa'\kappa}{\sqrt{2}}\frac{\upsilon_{w}}{M_{F}},\\
&g_{Z}=-\frac{\kappa g}{2\sqrt{2}\cos\theta_{w}}\frac{\upsilon_{w}}{M_{F}},\\
&g_{W}=\frac{\kappa g}{2}\frac{\upsilon_{w}}{M_{F}}.
\end{split}
\end{eqnarray}
Where $M_{F}$ is regarded as the common mass of the VLLs of all flavors, and $\upsilon_{w}\simeq$ 246 GeV is the electroweak symmetry breaking scale.

Then the partial  decay widths of the singlet VLL to SM final states can be written as
\begin{eqnarray}
\begin{split}
&\Gamma(\psi \rightarrow W\nu)=g_{W}^{2}\frac{M_{F}}{32\pi}(1-r_{W}^{2})^{2}(2+1/r_{W}^{2}) , \\
&\Gamma(\psi \rightarrow Z\ell)=g_{Z}^{2}\frac{M_{F}}{32\pi}(1-r_{Z}^{2})^{2}(2+1/r_{Z}^{2}),\\
&\Gamma(\psi \rightarrow h\ell)=\kappa^{2}\frac{M_{F}}{64\pi}(1-r_{h}^{2})^{2},\\
\end{split}
\end{eqnarray}
where $r_{X}=M_{X}/M_{F}$ for $X= W, Z, h$. Certainly, if the singlet VLL mass is larger than the mass of the new scalar $S$, it is possible that the singlet VLL decays into $S$ and a SM charged lepton. However, in this paper, we assume that the new scalar $S$ is generally heavier than the singlet VLL, and thus do not consider this decay channel.

\begin{figure}[H]
\begin{center}
\subfigure[]{\includegraphics [scale=0.6] {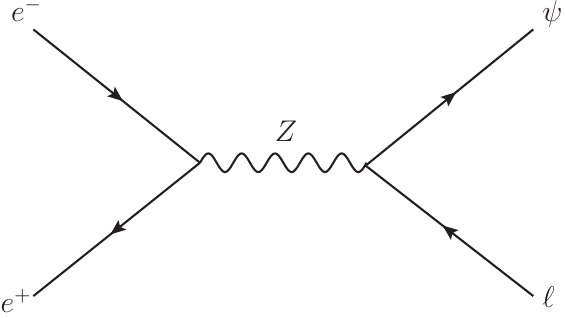}}
\hspace{0.4in}
\subfigure[]{\includegraphics [scale=0.6] {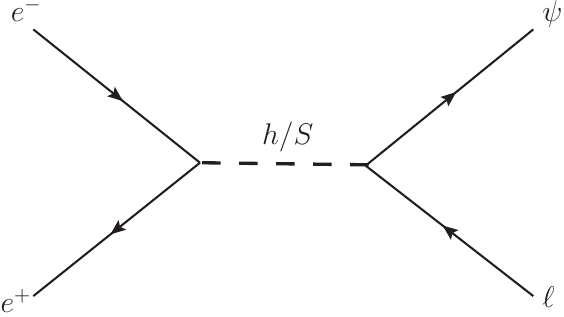}}
\hspace{0.4in}
\subfigure[]{\includegraphics [scale=0.6] {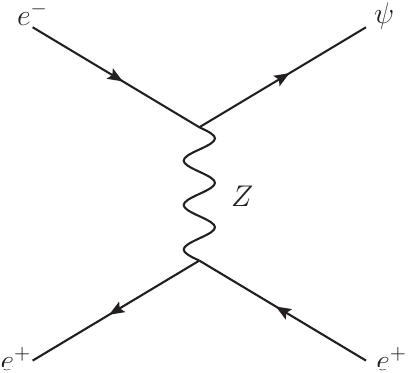}}
\hspace{1.0in}
\subfigure[]{\includegraphics [scale=0.6] {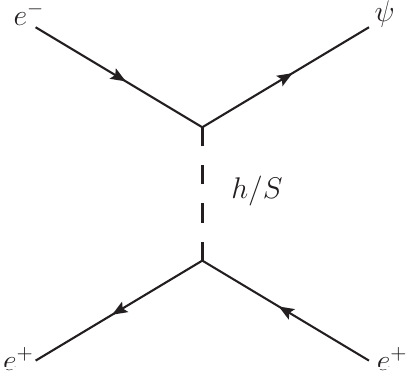}}
\caption{Feynman diagrams for single production of VLL at the ILC.}
\label{feynman diagram}
\end{center}
\end{figure}
From Eq.(1) we can see that VLL can be single produced at the ILC via the process $e^{+}e^{-}\rightarrow\psi \ell$ mediated by the gauge boson $Z$ or the scalars $S$ and $H$. The relevant Feynman diagrams are shown in Fig.~\ref{feynman diagram}. However, the main contributions come from exchange of the gauge boson $Z$. Furthermore,  the t-channel contributions are smaller than those of the s-channel and only contribute to the final state $\psi e$ in the case of flavor conservation. All the same, we will consider all of these contributions in our numerical estimations.

If the lepton mass of the final state is disregarded, then the production cross sections of the processes $e^{+}e^{-}\rightarrow$ $\psi e$, $\psi\mu$ and $\psi\tau$ are approximately equal to each other. Nevertheless, for $\tau$ lepton, it decays more rapidly and the background is more intricate. For simplicity, we take the first generation of leptons as an example to conduct numerical calculation and the analysis of the signal and SM background in this paper. From above discussions we can see that the production cross sections are contingent upon the free parameters ${M_{F}}, {M_{S}}, \kappa'$ and $ \kappa$. For better clarity, we give the expression form of the production cross section of the s-channel process $e^{+}e^{-}\rightarrow$ $\psi e$ only induced by exchange of the gauge boson $Z$

\begin{eqnarray}
\begin{split}
\sigma(e^{+}e^{-}\rightarrow\psi~e)
    = \frac{e^2\kappa^2\upsilon_{w}^2(8\sin^4\theta_{w}-4\sin^2\theta_{w}+1)(s-M_F^{2})^2(M_F+2s)}{3072\pi \sin^4\theta_{w}\cos^4\theta_{w}s^{2}M_F(s-M_Z^{2})}.
\end{split}
\end{eqnarray}

The constraints on the free parameters of the VLS models have been extensively studied in Refs.~\cite{Hiller:2020fbu,Hiller:2019mou,Bissmann:2020lge,Yue:2024sds}, which mainly come from measurements of the muon and electron AMMs and the relevant electroweak data. The contributions of  the VLS models to the muon AMM, $a_{\mu}$, are given by~\cite{Hiller:2020fbu,Hiller:2019mou}
\begin{eqnarray}
\begin{split}
&\Delta{a_{\mu} = \frac{\kappa'^2}{32\pi^{2}} \frac{m_{\mu}^{2}}{M_{F}^{2}} f(\frac{M_{S}^{2}}{M_{F}^{2}} )}
\end{split}
\end{eqnarray}
with $f(t) = (2t^{3} + 3t^{2} - 6t^{2}\ln{t} - 6t + 1)/ (t - 1)^{4}$ positive for any $t$, and $f(0)= 1$. If we demand the VLS models to solve the muon AMM anomaly,  then the coupling $\kappa'$ can be expressed in terms of the mass parameters $M_{F}$ and $M_{S}$.  Similar to $a_{\mu}$, the contributions of the VLS models to the electron AMM, $a_{e}$, is related to the coupling $\kappa$, the mass parameters $M_{F}$ and $M_{S}$. According to both AMMs and electroweak data, we simply fix $\kappa/\kappa' = 10^{-2}$ with $\kappa' $ computed from Eq.(6) for $M_{S}=$ 500 GeV and $\kappa\in$ (0.01,~0.1) in our following numerical calculation.

\section{Searching for singlet VLL via its single production at the ILC}
\begin{figure}[H]
\begin{center}
\centering\includegraphics [scale=0.60] {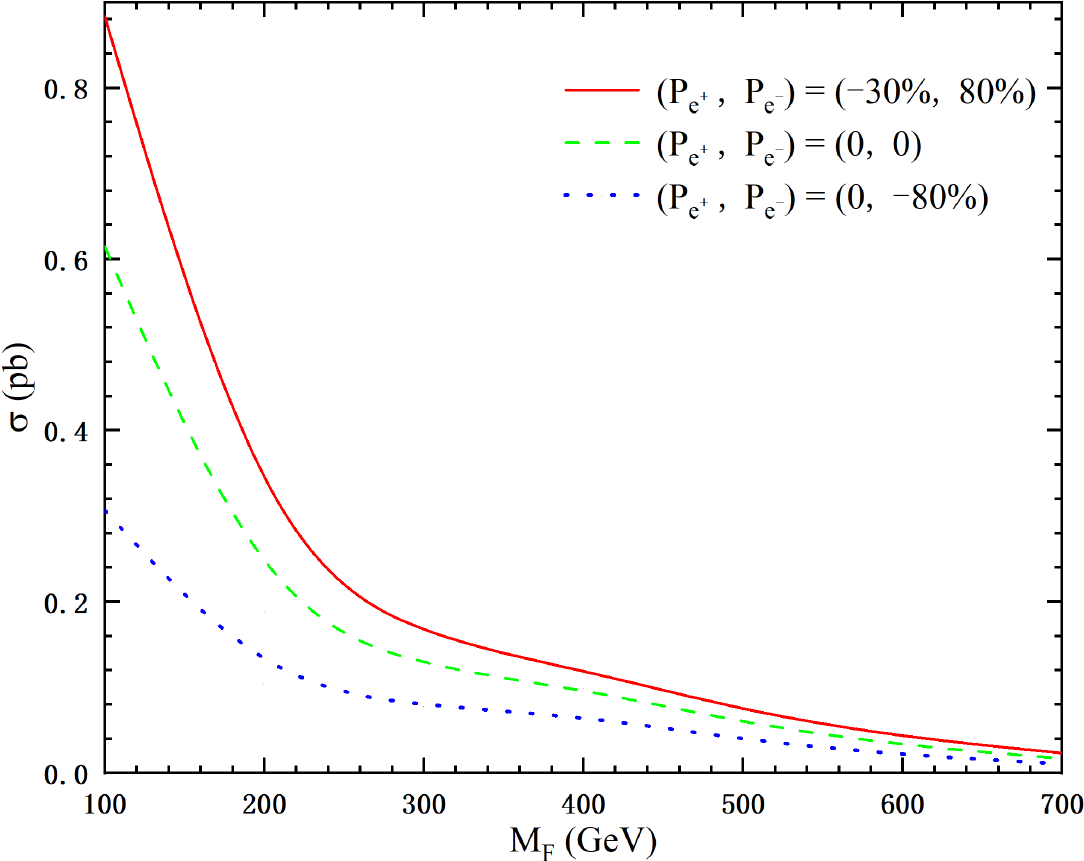}
\caption{Production cross section of the process $e^{+}e^{-}\rightarrow\psi e^{+}$ as a function of the mass parameter $M_{F}$ for different polarization options with the coupling $\kappa = $ 0.03.}
\label{cross1}
\end{center}
\end{figure}

 As polarized $e^{+}$ and $e^{-}$ beams can change the production cross sections, in Fig.~\ref{cross1}, we plot the production cross section of the process $e^{+}e^{-}\rightarrow\psi e^{+}$ with different polarization options as a function of the mass parameter $M_{F}$ at the 1 TeV ILC for the coupling parameter $\kappa = $ 0.03. It can be observed from Fig.~\ref{cross1} that the values of the production cross section  attains maximum when the polarization is $-$30\% for positron beam and 80\% for electron beam, which are in the range of 0.023 - 0.883 pb for 100 GeV$ <M_{F}<$ 700 GeV. In order to produce a heavier VLL in a larger parameter space of the singlet VLS model, we choose polarized beams with $P(e^+, e^-)= (-30\%,~80\%)$ in the following analysis of the signal and SM background.

According to the discussions given in Ref.~\cite{Bissmann:2020lge} about the branching ratios for various possible decay channels of singlet VLL and the decay formulas given above, we find that the value of the branching ratio $Br(\psi \rightarrow W\nu)$ is the largest. So, we prefer to choose the most outstanding decay channel for our analysis. Furthermore, in the following analysis of the signal and SM background, we assume that W boson decays through two kinds of mode, that are pure leptonic  and fully hadronic decay channels. The production and decay chain are presented as follows:
\\(1) pure leptonic decay channel:~~~~~~~~~$e^{+}e^{-}\rightarrow\psi e^{+}\rightarrow W^{-}\nu_{e}e^{+}\rightarrow \ell^{-}\bar{\nu_{\ell}}\nu_{e}e^{+}$;
\\(2) fully hadronic decay channel:~~~~~~~$e^{+}e^{-}\rightarrow\psi e^{+}\rightarrow W^{-}\nu_{e}e^{+}\rightarrow jj\nu_{e}e^{+}$.

For further calculation, we use \verb"FeynRules" package~\cite{Alloul:2013bka} to generate the model file and export it to the \verb"UFO" format~\cite{Degrande:2011ua}. The simulation of the signal and SM background at leading order (LO) is performed with \verb"MadGraph5_aMC@NLO"~\cite{Alwall:2014hca,Frederix:2018nkq}. The basic cuts we use first are given by
$$
\begin{array}{l}
	~~~~~~\textit{p}_{T}^{\ell}>10 \mathrm{GeV}, ~~\quad  \lvert \eta_{\ell}\rvert<2.5,\\
~~~~~~\textit{p}_{T}^{j}>20 \mathrm{GeV}, ~~\quad  \lvert \eta_{j}\rvert<5.0,\\
    ~\quad  \Delta R(x,~y)>0.4,~with~x, y = \ell,~j(u,d,c,s).\\
\end{array}
$$
$\textit{p}_{T}^{\ell}$ and $\textit{p}_{T}^{j}$ are the transverse momentum of leptons and jets. $\lvert \eta_{\ell}\rvert$ and $\lvert \eta_{j}\rvert$ are the pseudo-rapidity of leptons and jets. $\Delta R(x,~y)$ is defined as the separation in the rapidity-azimuth plane, $\Delta R(x,y)=\sqrt{(\Delta\phi)^{2}+(\Delta\eta)^{2}}$ with $\phi$ being azimuthal angle. Then we use \verb"PYTHIA8"~\cite{Sjostrand:2014zea} to simulate initial state radiation, final state radiation and hadronization. The fast detector simulation is made by \verb"DELPHES 3"~\cite{deFavereau:2013fsa} using the International Linear Detector (ILD) detector card. To realize the kinematic and cut-based analysis, the package \verb"MADANALYSIS 5"~\cite{Conte:2012fm,Conte:2014zja,Conte:2018vmg} is adopted.


\subsection{Pure leptonic decay channel}

For the pure leptonic decay channel, the final state  is composed of  two charged leptons and missing energy. The corresponding SM background mainly comes from the processes $e^{+}e^{-}\rightarrow\ell^{+}\ell^{-}\nu_{\ell}\bar{\nu_{\ell}}$ ~and~ $e^{+}e^{-}\rightarrow\ell^{+}\ell^{-}\nu_{\ell}\nu_{\ell}\bar{\nu_{\ell}}\bar{\nu_{\ell}}$ with $\ell=e,~\mu$. The possible subprocesses that lead to the similar final states comprise of ~$e^{+}e^{-}\rightarrow W^{+}W^{-}\rightarrow\ell^{+}\ell^{-}\nu_{\ell}\bar{\nu_{\ell}}$, ~$e^{+}e^{-}\rightarrow ZZ\rightarrow\ell^{+}\ell^{-}\nu_{\ell}\bar{\nu_{\ell}}$, ~$e^{+}e^{-}\rightarrow W^{+}W^{-}Z\rightarrow\ell^{+}\ell^{-}\nu_{\ell}\nu_{\ell}\bar{\nu_{\ell}}\bar{\nu_{\ell}}$ ~and~ $e^{+}e^{-}\rightarrow ZZZ\rightarrow\ell^{+}\ell^{-}\nu_{\ell}\nu_{\ell}\bar{\nu_{\ell}}\bar{\nu_{\ell}}$. For the SM background, the value of the production cross section is about 0.08931 pb. As for the signal, the values of the production cross section can respectively reach $3.8301\times10^{-7}$ pb, $5.2597\times10^{-8}$ pb  and $9.8058\times10^{-9}$ pb for   $M_{F} = $ 600 GeV, 700 GeV and 800 GeV. Considering collision energy and the lower limits of current experiments on the masses of the singlet VLLs, we take the VLL mass as $M_F\in$ (300,~700) GeV.

\begin{figure}[H]
\begin{center}
\subfigure[]{\includegraphics [scale=0.5] {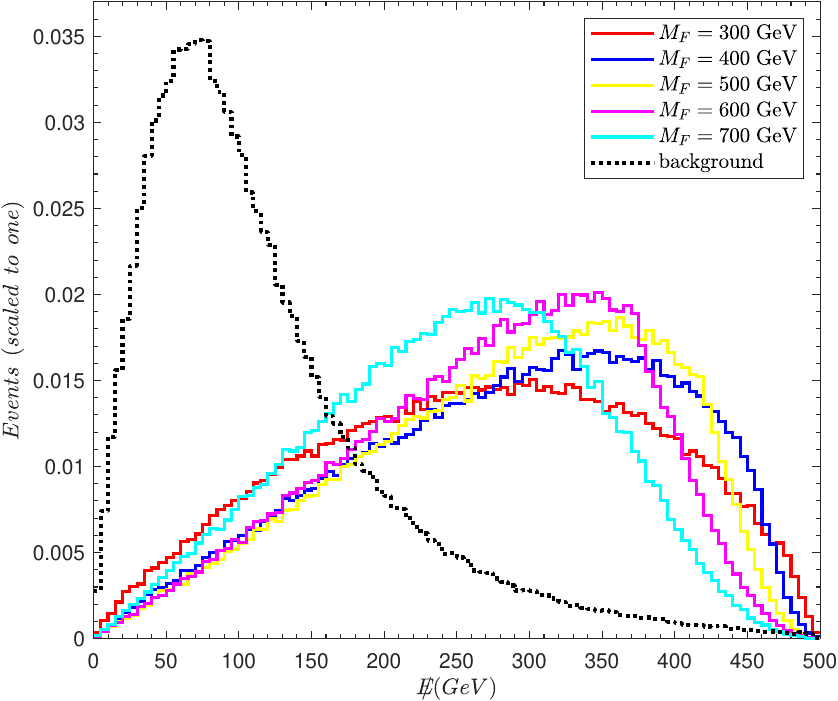}}
\hspace{0.2in}
\subfigure[]{\includegraphics [scale=0.5] {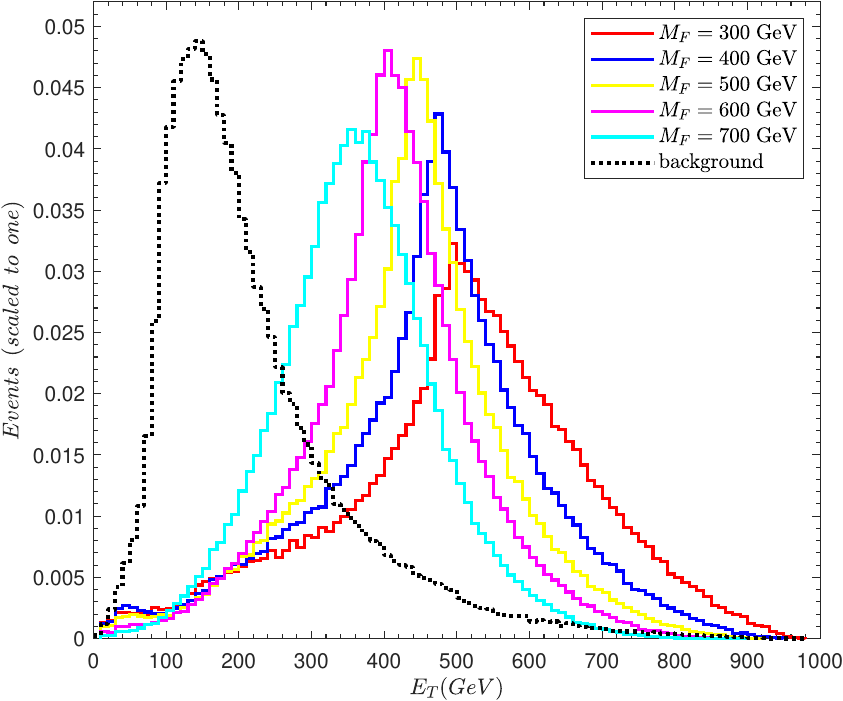}}
\hspace{0.8in}
\subfigure[]{\includegraphics [scale=0.5] {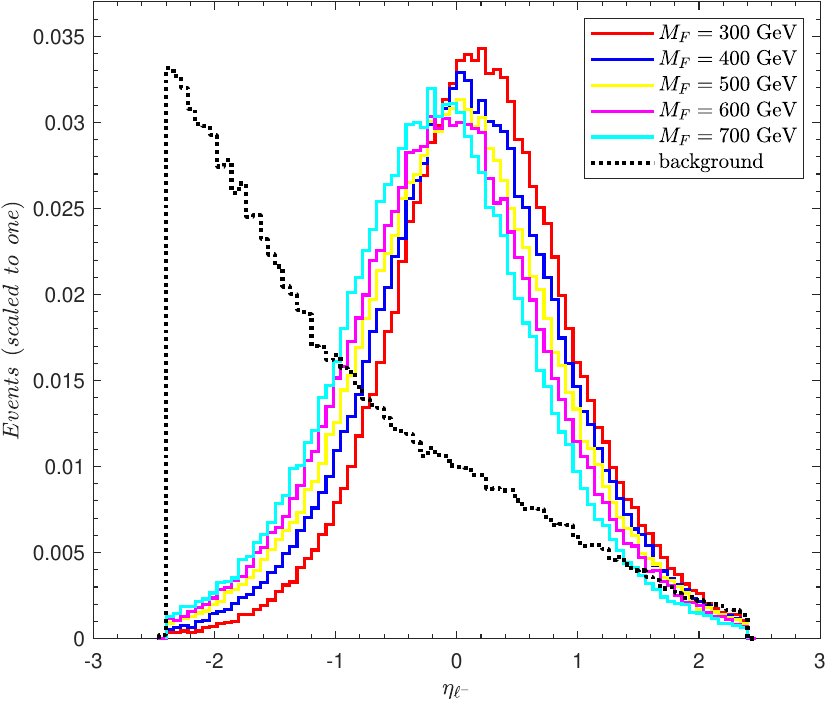}}
\hspace{0.2in}
\subfigure[]{\includegraphics [scale=0.5] {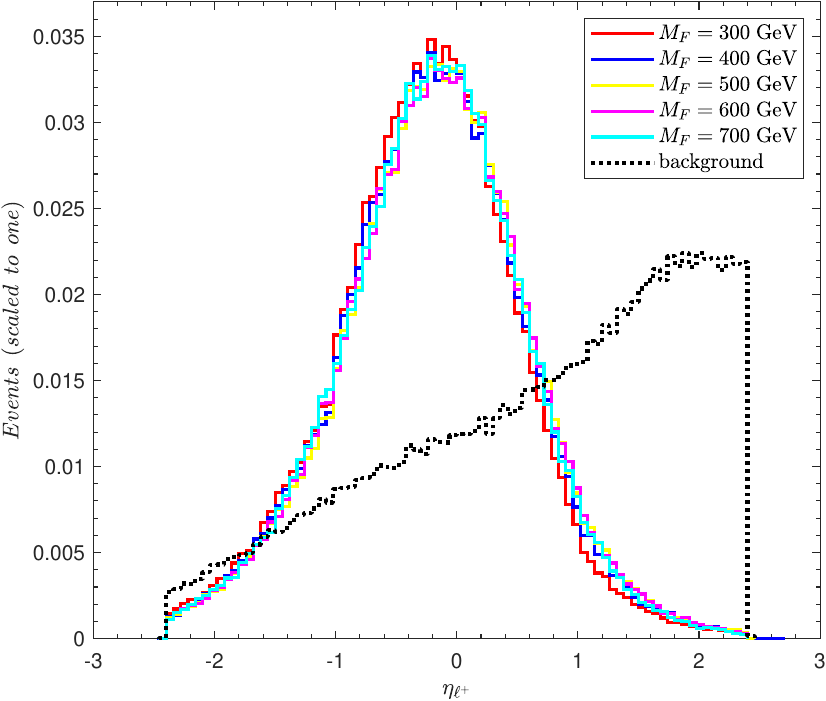}}
\caption{The normalized distributions of $E\mkern-10.5 mu/$ (a), $E_T$ (b), $\eta_{\ell^{-}}$ (c) and $\eta_{\ell^{+}}$ (d) of signal and background for various benchmarks in pure leptonic decay channel at the 1 TeV ILC with $\mathcal{L}=$ 1~ab$^{-1}$. The solid and dashed lines stand for the signal and background events, respectively.}
\label{distribution1}
\end{center}
\end{figure}

We perform a cut-based analysis by looking at some relevant kinematic variables. In order to improve the signal-to-noise ratio, appropriate cuts are applied to these variables. The normalized distributions of $E\mkern-10.5 mu/$, ~$E_T$, ~$\eta_{\ell^{-}}$, ~$\eta_{\ell^{+}}$ are presented in Fig.~\ref{distribution1}, where the solid lines stand for signal events and dashed lines stand for background events. The benchmark points $M_{F} = $ 300 GeV, 400 GeV, 500 GeV, 600 GeV and 700 GeV are selected with the assumption of $\kappa =$ 0.03.

$E\mkern-10.5 mu/$ denotes missing transverse energy and its distribution is shown in Fig.~\ref{distribution1}(a). For the SM background, the distribution peaks at lower values of $E\mkern-10.5 mu/$, while for signal it is distributed at higher values. The reason is that the missing particles $\nu_{\ell}$ decay from the heavier mother particle $\psi$ for the signal events, so more mass turns into the momentum of missing particles. A lower cut of $E\mkern-10.5 mu/ >$ 100 GeV helps to diminish the SM background.

$E_T$ presents scalar sum of the transverse energy of all final-state particles. From Fig.~\ref{distribution1}(b), we can see that the signal events with different $M_{F}$ are almost distributed in the region with around 400 GeV and the background events are distributed at around 150 GeV. So we take the cut $E_T > $ 200 GeV to suppress the SM background peak.

The normalized pseudorapidity distribution of the charged lepton $\ell^{-}$, $\ell^{+}$ are depicted in Fig.~\ref{distribution1}(c) and (d). As shown in these pictures, the SM background has a peak at low values for $\eta_{\ell^{-}}$, while the SM background has a peak at high values for $\eta_{\ell^{+}}$. It is because the leptons can produced via s-channel and t-channel exchange of $\gamma$, $Z$ and $\nu$. Whereas, as for the signal, the leading and sub-leading leptons are mainly produced via s-channel processes. So, the $\eta$ distribution for the signal is more centrally peaked. Thus, we choose $\eta_{\ell^{-}}>$ $-$ 1.0 and $\eta_{\ell^{+}}<$ $+$ 1.0 to reduce the background significantly.

\begin{table}[H]
\begin{center}
\caption{All cuts on the signal and SM background for pure leptonic decay channel.}
\label{tablecut1}
\begin{tabular}
[c]{c| c c c c c}
    \hline\hline
    ~~~~Basic cuts~~~~      &  $\textit{p}_{T}^{\ell}>10 \mathrm{GeV}, ~~ \quad  \lvert \eta_{\ell}\rvert<2.5, ~~ \quad  \textit{p}_{T}^{j}>20 \mathrm{GeV},$\\&~~$\quad  \lvert \eta_{j}\rvert<5.0,~\quad  \Delta R(x,~y)>0.4,~with~x, y = \ell,~j~~~$ \\

	~~~~Cut 1~~~~      &  $~~~~~~~~~~~~~~~~~E\mkern-10.5 mu/>100~GeV~~~~~~~~~~~~~~$ \\

	~~~~Cut 2~~~~      &  $~~~~~~~~~~~~~~~~~E_T>200~GeV~~~~~~~~~~~~~~$  \\

	~~~~Cut 3~~~~      &  $~~~~~~~~~~~~~~~~~\eta_{\ell^{-}}> -1.0~~~~~~~~~~~~~~$    \\

    ~~~~Cut 4~~~~      &  $~~~~~~~~~~~~~~~~~\eta_{\ell^{+}}<~  1.0~~~~~~~~~~~~~$    \\
   \hline \hline

\end{tabular}
\end{center}
\end{table}

The optimized selection cuts based on the characteristics of the kinematics distributions are listed in Table~\ref{tablecut1}. In Table~\ref{tablecs1}, we tabulate the cross sections of the signal and SM background surviving after the application of each cut for five benchmarks. The statistical significance (SS) is evaluated by the formula $SS=S/\sqrt{S+B}$ for an integrated luminosity of 1 ab$^{-1}$, where $S$ and $B$ are the number of events for the signal and background, respectively. From Table~\ref{tablecs1} we infer that, to attain 5$\sigma$ significance, the mass of VLL must be less than 400 GeV. The SS can reach 6.282 for $M_{F}=$ 300 GeV and $\kappa =$ 0.03.

\begin{table}[H]\tiny
	\centering{
\caption{The cross sections of the signal and SM background processes after the improved cuts applied for $\kappa=0.03$ at the 1TeV ILC with benchmark points for pure leptonic decay channel.$~$\label{tablecs1}}
		\newcolumntype{C}[1]{>{\centering\let\newline\\\arraybackslash\hspace{50pt}}m{#1}}
		\begin{tabular}{m{1.5cm}<{\centering}|m{2cm}<{\centering} m{2cm}<{\centering} m{2cm}<{\centering}  m{2cm}<{\centering} m{2cm}<{\centering} m{2cm}<{\centering}}
			\hline \hline
      \multirow{2}{*}{Cuts} & \multicolumn{5}{c}{cross sections for signal (background) [pb]}\\
     \cline{2-6}
     & $M_F=300$ GeV  & $M_F=400$ GeV  & $M_F=500$ GeV & $M_F=600$ GeV &  $M_F=700$ GeV  \\ \hline
     Basic Cuts  & \makecell{$7.9396\times10^{-4}$\\$(0.08931)$} & \makecell{$4.4850\times10^{-4}$\\$(0.08931)$} &\makecell{$6.0852\times10^{-5}$\\$(0.08931)$} &\makecell{$3.8301\times10^{-7}$\\$(0.08931)$} & \makecell{$5.2597\times10^{-8}$\\$(0.08931)$}
        \\
     Cut 1  & \makecell{$7.4610\times10^{-4}$\\$(0.04272)$} & \makecell{$4.2487\times10^{-4}$\\$(0.04272)$} &\makecell{$5.7407\times10^{-5}$\\$(0.04272)$} &\makecell{$3.5401\times10^{-7}$\\$(0.04272)$} & \makecell{$4.7297\times10^{-8}$\\$(0.04272)$}
       \\
     Cut 2  &\makecell{$7.0624\times10^{-4}$\\$(0.02505)$}  & \makecell{$4.0523\times10^{-4}$\\$(0.02505)$} &\makecell{$5.5127\times10^{-5}$\\$(0.02505)$} &\makecell{$3.3601\times10^{-7}$\\$(0.02505)$} & \makecell{$4.2698\times10^{-8}$\\$(0.02505)$}
       \\
     Cut 3  & \makecell{$5.6228\times10^{-4}$\\$(0.01216)$} & \makecell{$3.1377\times10^{-4}$\\$(0.01216)$} &\makecell{$4.1767\times10^{-5}$\\$(0.01216)$} &\makecell{$2.5301\times10^{-7}$\\$(0.01216)$} & \makecell{$3.2698\times10^{-8}$\\$(0.01216)$}
     \\
     Cut 4  & \makecell{$5.1152\times10^{-4}$\\$(6.1029\times10^{-3})$} & \makecell{$2.8332\times10^{-4}$\\$(6.1029\times10^{-3})$} &\makecell{$3.7550\times10^{-5}$\\$(6.1029\times10^{-3})$} &\makecell{$2.2701\times10^{-7}$\\$(6.1029\times10^{-3})$} & \makecell{$2.9498\times10^{-8}$\\$(6.1029\times10^{-3})$}
     \\ \hline
     $SS$  & $6.282$ & $3.541$ & $0.479$ & $0.00291$ & $0.000378$  \\ \hline \hline
	\end{tabular}}	
\end{table}

\subsection{Fully hadronic decay channel}
In this subsection, we investigate the signal induced by  the boson W decaying into hadrons, utilizing an analysis method similar to the pure leptonic decay channel. The SM background is dominantly from the process $e^{+}e^{-}\rightarrow jj\ell^{+}\nu_{\ell}$. In order to effectively distinguish the signal from the SM background, we select the cuts on two important kinematic variables: $M_{eff}$, $E_T$.

\begin{figure}[H]
\begin{center}
\subfigure[]{\includegraphics [scale=0.5] {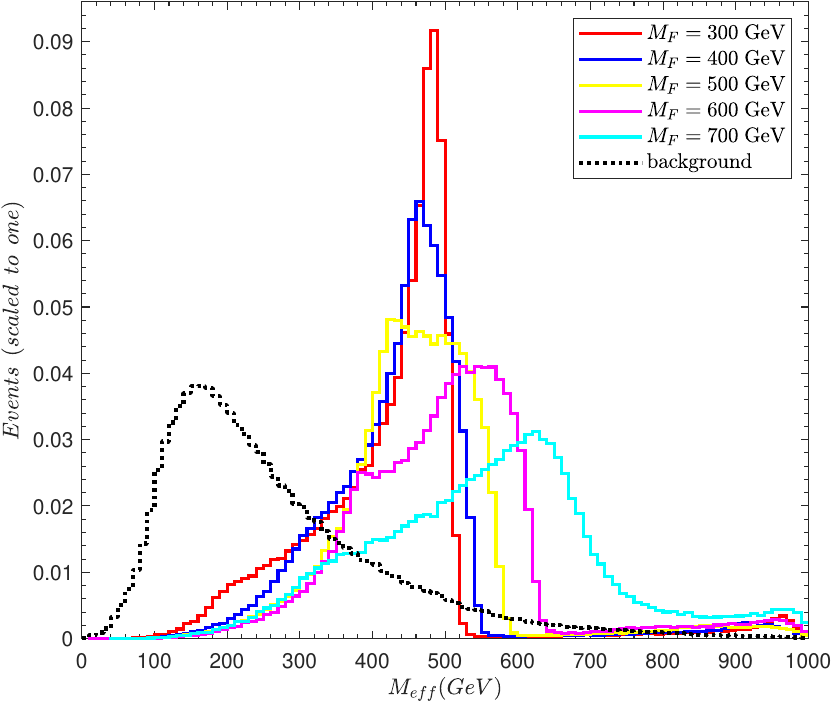}}
\hspace{0.2in}
\subfigure[]{\includegraphics [scale=0.5] {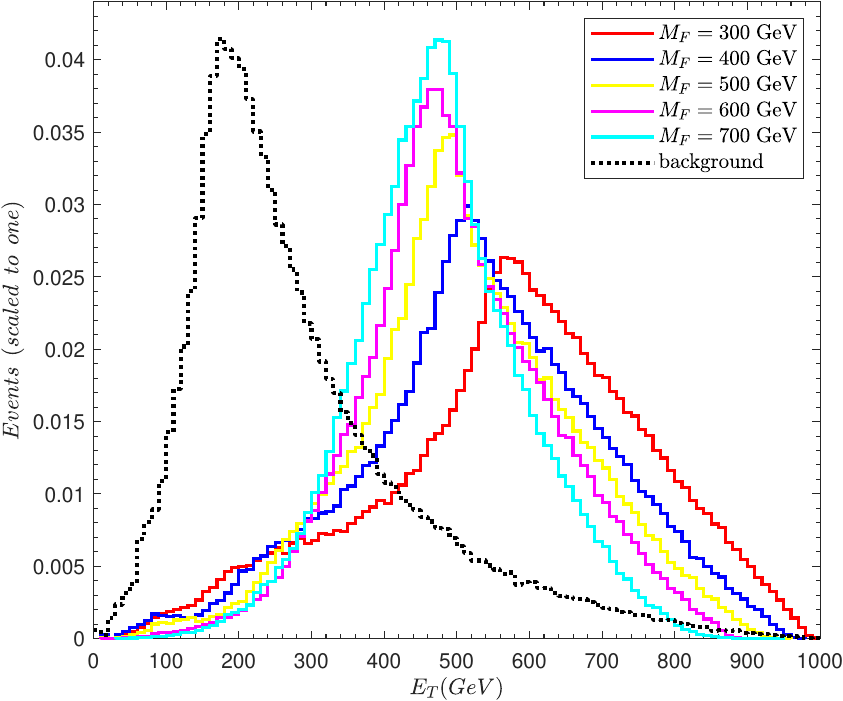}}
\caption{The normalized distributions of $M_{eff}$ (a) and $E_T$ (b) for fully hadronic decay channel at the 1~TeV ILC with $\mathcal{L}=$ 1 ab$^{-1}$.}
\label{distribution2}
\end{center}
\end{figure}

$M_{eff}$ is defined as the scalar sum of the lepton transverse momentum $\textit{p}_{T}$ and the missing energy $E\mkern-10.5 mu/$. We find that putting a cut on $M_{eff}$ is more effective for rejecting background than giving cuts on $\textit{p}_{T}$ and $E\mkern-10.5 mu/$ separately. From the distribution shown in Fig.~\ref{distribution2}(a), we impose a cut $M_{eff}>$ 200 GeV to improve the signal over the background.

From Fig.~\ref{distribution2}(b), it can be seen that the peak distribution of $E_T$ for the signal of fully hadronic decay channel is between 500 - 600 GeV, which is shifted to the right compared to the peak distribution of the pure leptonic decay channel given in Fig.~\ref{distribution1}(b), which is between 400 - 500 GeV. This is mainly because there is only one missing particle in fully hadronic decay channel, which makes the sum of the transverse momenta of the visible particles larger than that in pure leptonic decay channel. For the SM background, the distribution peaks at around 170 GeV, while for the signal it peaks at 500 GeV approximately. So we take $E_T>$ 300 GeV to enhance the signal significance.

\begin{table}[H]
\begin{center}
\caption{All cuts on the signal and SM background for fully hadronic decay channel.}
\label{tablecut2}
\begin{tabular}
[c]{c| c c c c c}
    \hline\hline
    ~~~~Basic cuts~~~~      &  $\textit{p}_{T}^{\ell}>10 \mathrm{GeV}, ~~ \quad  \lvert \eta_{\ell}\rvert<2.5, ~~ \quad  \textit{p}_{T}^{j}>20 \mathrm{GeV},$\\&~~$\quad  \lvert \eta_{j}\rvert<5.0,~\quad  \Delta R(x,~y)>0.4,~with~x, y = \ell,~j~~~$ \\

	~~~~Cut 1~~~~      &  $~~~~~~~~~~~~~~~~~M_{eff}>200~GeV~~~~~~~~~~~~~~$ \\

	~~~~Cut 2~~~~      &  $~~~~~~~~~~~~~~~~~E_T>300~GeV~~~~~~~~~~~~~~$  \\
   \hline \hline

\end{tabular}
\end{center}
\end{table}

All selected cuts on fully decay hadronic channel are organized in Table~\ref{tablecut2}. The values of the significance $SS$ are summarized in the last column of Table~\ref{tablecs2} for benchmark points. It is obvious that this decay channel performs better than the pure leptonic decay channel. The value of $SS$ can reach 7.65 for $M_{F}=$ 400 GeV and the integrated luminosity $\mathcal{L}$ = 1 ab$^{-1}$. Certainly, the value of $SS$ decreases as  the singlet VLL mass $M_{F}$ increasing, and its value becomes very small for $M_{F}\geq$ 600 GeV.

\begin{table}[H]\tiny
	\centering{
\caption{The cross sections of the signal and background processes after the improved cuts applied for $\kappa=0.03$ at the 1TeV ILC with benchmark points for fully hadronic decay channel.$~$\label{tablecs2}}
		\newcolumntype{C}[1]{>{\centering\let\newline\\\arraybackslash\hspace{50pt}}m{#1}}
		\begin{tabular}{m{1.5cm}<{\centering}|m{2cm}<{\centering} m{2cm}<{\centering} m{2cm}<{\centering}  m{2cm}<{\centering} m{2cm}<{\centering} m{2cm}<{\centering}}
			\hline \hline
      \multirow{2}{*}{Cuts} & \multicolumn{5}{c}{cross sections for signal (background) [pb]}\\
     \cline{2-6}
     & $M_F=300$ GeV  & $M_F=400$ GeV  & $M_F=500$ GeV & $M_F=600$ GeV &  $M_F=700$ GeV  \\ \hline
     Basic Cuts  & \makecell{$2.3803\times10^{-3}$\\$(0.07540)$} & \makecell{$1.3456\times10^{-3}$\\$(0.07540)$} &\makecell{$1.8250\times10^{-4}$\\$(0.07540)$} &\makecell{$1.1485\times10^{-6}$\\$(0.07540)$} & \makecell{$1.5774\times10^{-7}$\\$(0.07540)$}
        \\
     Cut 1  & \makecell{$2.3540\times10^{-3}$\\$(0.04458)$} & \makecell{$1.3362\times10^{-3}$\\$(0.04458)$} &\makecell{$1.8192\times10^{-4}$\\$(0.04458)$} &\makecell{$1.1383\times10^{-6}$\\$(0.04458)$} & \makecell{$1.5554\times10^{-7}$\\$(0.04458)$}
       \\
     Cut 2  &\makecell{$2.1566\times10^{-3}$\\$(0.02510)$}  & \makecell{$1.2427\times10^{-3}$\\$(0.02510)$} &\makecell{$1.7271\times10^{-4}$\\$(0.02510)$} &\makecell{$1.0786\times10^{-6}$\\$(0.02510)$} & \makecell{$1.4476\times10^{-7}$\\$(0.02510)$}
     \\ \hline
     $SS$  & $13.06$ & $7.6531$ & $1.0834$ & $0.00683$ & $0.000916$  \\ \hline \hline
	\end{tabular}}	
\end{table}

\begin{figure}[H]
\begin{center}
\subfigure[]{\includegraphics [scale=0.35] {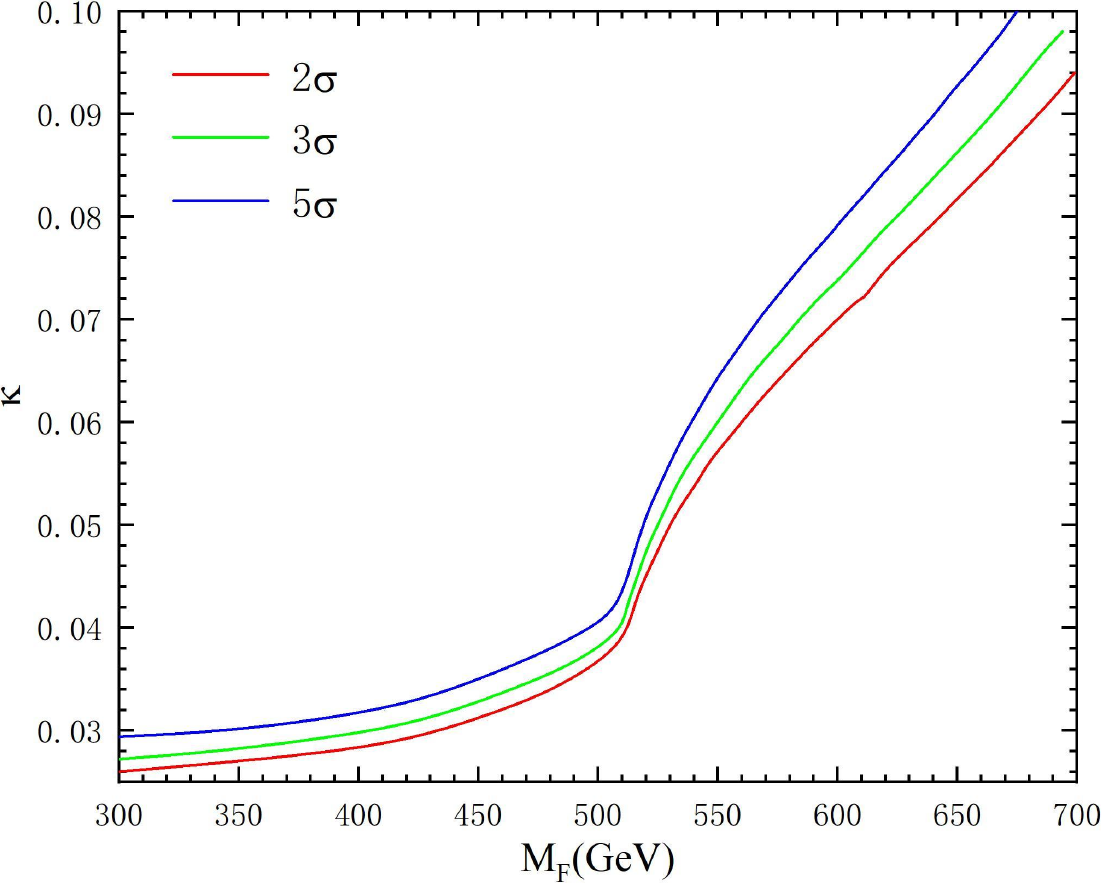}}
\hspace{0.1in}
\subfigure[]{\includegraphics [scale=0.35] {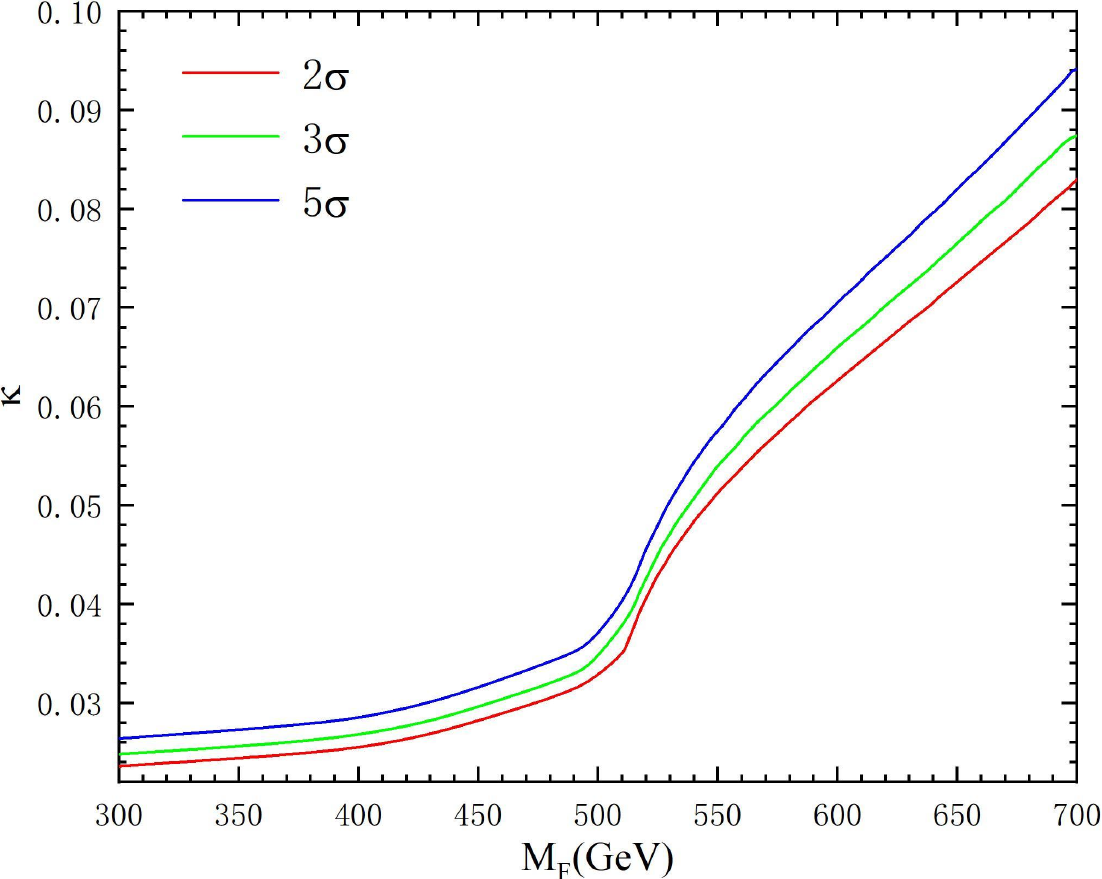}}
\caption{The 2$\sigma$, 3$\sigma$ and 5$\sigma$ curves in the $M _F - \kappa$ plane for pure leptonic (left) and fully hadronic (right) decay channels at the 1 TeV ILC with $\mathcal{L}=$ 1 ab$^{-1}$ .}
\label{235sigma}
\end{center}
\end{figure}

In Fig.~\ref{235sigma}, we plot the 2$\sigma$, 3$\sigma$ and 5$\sigma$ curves for the 1 TeV ILC with $\mathcal{L}$ = 1 ab$^{-1}$ in the plane ($M_F$, $\kappa$). As shown in Fig.~\ref{235sigma}, the signal of singlet VLL might be probed via the pure leptonic and fully hadronic decay channels. However, the latter is more sensitive to search for singlet VLL in the studied mass range via its single production at the 1 TeV ILC with $\mathcal{L}$ = 1 ab$^{-1}$. The prospective sensitivities to the coupling coefficient $\kappa$ can be as low as 0.0260, 0.0272, 0.0294 and 0.0236, 0.0248,~0.0264 at 2$\sigma$, 3$\sigma$, 5$\sigma$ levels for pure leptonic and fully hadronic decay channels, respectively. Thus, we can say that the future lepton collider ILC has the potential for detecting  singlet VLLs in most of the mass range considered in this paper.

\section{Conclusions}

So far, a large number of new physics models have been proposed to solve the SM problems and the existed anomalies, which predict the existence of new particles. Searching for the possible physics signals of various new particles are the main goals of current and future collider experiments. VLLs as one kind of  interesting new particles can produce rich phenomenology at low- and high-energy experiments, which should be extensively studied.

In this paper, we investigate the possibility of detecting singlet VLL through its single production at the 1 TeV ILC with $\mathcal{L}$ = 1 ab$^{-1}$ and the polarization option $P(e^{+}, e^{-})=(-30\%, 80\%)$. For the signal and SM background analysis, we have considered two kinds of decay channels for the W boson, i.e. pure leptonic and fully hadronic decay channels, which  produce two kinds of the signals: one is two charged leptons plus missing energy, the other is two jets, one charged lepton plus missing energy. Our analytic results for the first and second generations of singlet VLLs show that the parameter space $M_{F}\in$ [300, 675] GeV and $\kappa \in$ [0.0294, 0.1] might be detected by the proposed ILC for pure leptonic decay channel. For fully hadronic decay channel, larger detection regions of $M_{F}\in$ [300, 700] GeV and $\kappa \in$ [0.0264, 0.0941] are derived. Comparing the results of the two cases, we can see that singlet VLL might be more easily detected at the ILC via its single production in fully hadronic decay than that in pure leptonic decay channel.

\section*{ACKNOWLEDGMENT}

This work was partially supported by the National Natural Science Foundation of China under Grant No. 11875157, No. 12147214 and No. 11905093.


\bibliography{scriptbib}

\end{document}